# Possible Experimental Evidence for Violation of Standard Electrodynamics, de Broglie Pilot Wave and Spacetime Deformation

R. Mignani[1], A. Petrucci[2], F. Cardone[3]

[1] Dipartimento di Fisica "E. Amaldi", Università degli Studi "Roma Tre", via della Vasca Navale 84, 00146 Roma, Italy +39 06 57337033 mignani@fis.uniroma3.it
[2] LNF – INFN via Enrico Fermi 40, 00044 Frascati (Rome) Italy
petruccia@fis.uniroma3.it
[3] Istituto per lo Studio dei Materiali Nanostrutturati (ISMN-CNR), via dei Taurini 6, 00185 Roma, Italy fabio.cardone@ismn.cnr.it

**Abstract**

**We report and discuss the results of double-slit-like experiments in the infrared range, which evidence an anomalous behaviour of photon systems under particular (energy and space) constraints. These outcomes apparently disagree both with standard quantum mechanics (Copenhagen interpretation) and with classical and quantum electrodynamics. Possible interpretations can be given in terms of either the existence of de Broglie–Bohm pilot waves associated to photons, and/or the breakdown of local Lorentz invariance (LLI). We put forward an intriguing hypothesis about the possible connection between these seemingly unrelated points of view by assuming that the pilot wave of a photon is, in the framework of LLI breakdown, a local deformation of the flat minkowskian spacetime.**

## 1. Introduction

The purpose of this paper is to present and discuss the outcomes obtained from optical experiments of the double-slit type which were designed in order to possibly shed some new light on two fundamental issues of Physics: wave-particle duality and the ranges of validity of local Lorentz Invariance. The former is a long-debated topic of Quantum Mechanics, but far from being resolved. The wave associated to a quantum object is commonly regarded as a probability wave, according to what is usually referred to as the Copenhagen interpretation, and hence it conveys no physical properties.

This interpretation is thoroughly antithetic to that advocated by Einstein, De Broglie and Bohm [1], which regards the quantum wave as real pilot wave, intimately bound to the quantum entity and moving along with it, but unable to carry either energy or momentum (hollow or ghost wave). Actually, all the experiments which evidence the wave–corpuscle duality require, in order to be correctly interpreted, the theoretical hypothesis of the objective reality of the quantum wave, but they do not provide any direct observation of the hollow wave. However, a way to detect such quantum waves might be through their affecting the probabilities of events to which they superimpose in space–time (for instance, in interference phenomena). This is essentially due (according to de Broglie and Andrade y Silva [2]) to the interaction of quantum objects with all the pilot waves present in a given space region, through their quantum potential [1].

The second fundamental issue discussed in this paper and addressed by our experiments has to do with the teaching of Einstein's relativity theories which state that physical phenomena occur in spacetime whose structure is globally curved (Riemannian) and locally flat (Minkowskian). The local flatness of spacetime means that the laws of physics can be locally written in the language of Special Relativity (SR) and hence physical phenomena are locally invariant under Lorentz transformations. The controversial point at issue (from both the theoretical and the experimental side) is whether the validity of local Lorentz invariance (LLI) is preserved at any length or energy scale. The experiments aimed at testing LLI can be roughly classified in three groups [3]: **i)** Michelson-Morley type experiments, which test isotropy of the round-trip speed of light; **ii)** experiments which test isotropy of the one-way speed of light, based on atomic spectroscopy and atomic timekeeping; **iii)** Hughes-Drever type experiments, testing isotropy of nuclear energy levels. All such experiments set upper bounds on the degree of violation of LLI. On the theoretical side, a lot of generalisations of Special Relativity and/or LLI breaking mechanisms exist in literature. For instance, in order to take account of the LLI breaking effects, an extension of the Standard Model has been proposed by Kostelecky [4], who puts very stringent limits by analysing the existing data. A possible signature for a breakdown of LLI in electromagnetic interactions was put forward by two of the present authors (F.C. and R.M.), in the framework of a generalisation of SR based on a "Local Deformation" of Space–Time (DST), assumed to be endowed with a metric whose coefficients depend on a parameter $E$ with the dimension of energy which is considered to be the energy exchanged during the non Lorentz invariant process[1] [5,6]. In order to determine the features of this parameter, the analytical forms of the metric coefficients and other phenomenological details of this theory, this formalism was used to analyse the experimental set-up and the results of two experiments carried out in Cologne [7] and Florence [8] where LLI was broken in the sense that superluminal propagation of electromagnetic waves was obtained. This analysis by the DST formalism produced a threshold value of energy[2] and space for the breakdown of LLI for the electromagnetic interaction. In particular LLI is broken when the energy exchanged during the process is less than 4.5 µeV and the maximum distance, over which its non Lorentzian effects can be still detected, is about 9 cm.

---

[1] It is necessary to clarify the meaning of the words "*a parameter with the dimension of energy which is considered to be the energy exchanged during the non Lorentz invariant process*". A physical process which is not invariant under Lorentz symmetry (in the local sense) takes places in or, more precisely, involves a locally non flat Minkowskian spacetime. In order to detect the effects due to a non flat spacetime (either curved or deformed) it is necessary to perform two measurements and then subtract their results. This is how Eddington operated when he measured the deflection of the light of a distant star around the Sun by subtracting two measured angles and this is how one operates with geodesic deviation to determine the intrinsic curvature of a manifold. In our case, the parameter $E$ parameterizes the local deformation of spacetime and, in this sense, has to be understood as the difference of two energies measured under two different conditions, as it happened for the angles in Eddington experiment.

The Cologne and Florence experiments, just mentioned, have to do with LLI breakdown from the point of view of superluminal propagation, which suggests the existence of a preferred inertial reference frame. However, a more comprehensive sight on LLI breakdown can be obtained from the point of view of the locally deformed spacetime (no more flat and rigid) which is a reasonable and quite evident consequence of the violation of this symmetry.

The design of the experiments, presented in this paper, finds its whys and wherefores in this last statement. First of all, we wanted to corroborate the hypothesis that LLI breakdown means indeed the existence of a locally deformed spacetime whose deformation stores some energy and is able to affect (pilot) the propagation of photons. The second part of the hypothesis establishes a connection between LLI breakdown and the Einstein-de Broglie-Bohm pilot wave. We hypothesise that the pilot wave is nothing but a deformation of spacetime geometry related to the quantum object (photon). Now, let's see what are the indications at our disposal for the design of the experiment, highlight some critical points and then describe the experimental set-up and the results obtained.

## 2. Experimental set-up Design and Results

Since we want to measure effects due to LLI breakdown we have to design the experimental set-up according to the energy and space threshold mentioned above. Moreover, since we are looking at LLI breakdown from the point of view of the effects brought about by the locally deformed spacetime on the propagation of photons, we have to search for these effects in a difference of two measurements according to what has been explained in the note 1 above. Besides, as it has been said, we want to shed new light on the de Broglie pilot wave and establish a connection between it and deformed spacetime. The presence of the de Broglie wave implies that the experiment will have to be designed with a set-up of double slit type. The information needed to accomplish these requirements is at our disposal and can be extracted from our phenomenological theory. However, there is a critical point which has to be faced in carrying out the experiments and is connected to the absence of any kind of indication on the time evolution of LLI breakdown. In other words, the effects that we want to search for by these experiments can have a very peculiar time evolution which is completely unknown and has to be experimentally taken into account and investigated. In the last decade we carried out three experiments involving photon systems in the near infrared range. Before moving on to the description we refer to Fig.1 where the lay-out of the set-up is reported. The distances between sources and detectors and in particular the distances L and s were fixed equal to those of the Florence experiment [8] where LLI breakdown showed up in the sense of superluminal propagation of electromagnetic waves in the microwave range between two horn antennas. A strong hypothesis was made here as to the independence of LLI breakdown from the range of frequency (microwave to infrared).

---

[2] In all but very few experiments, LLI is valid and the laws of physics are compliant with the language of special relativity. Hence, it goes without saying that there must exist an energy threshold either very low or very high (or maybe both) that acts as an upper bound or a lower bound respectively for those energy ranges where LLI is broken. In our phenomenological framework we search for a sufficiently low energy threshold below which LLI is violated.

## 2.1. Experimental set-up

The apparatus employed in all experiments (schematically depicted in Fig. 1) consisted of a Plexiglas box with wooden base and lid. The box (thoroughly screened from those frequencies susceptible of affecting the measurements) contained two identical infrared (IR) LEDs, as (incoherent) sources of light, and three identical detectors (A, B, C). The two sources S1, S2 were placed in front of a screen with three circular apertures F1, F2, F3 on it. The apertures F1 and F3 were lined up with the two LEDs A and C respectively, so that each IR beam propagated perpendicularly through each of them. The geometry of this equipment and the absorbing material of the internal walls were designed so that no photon could pass through aperture F2 on the screen. The wavelength of the two photon sources was $\lambda = 8.5 \times 10^{-5}$ cm. The apertures were circular, with a diameter of 0.5 cm, much larger than $\lambda$. We therefore worked in the absence of single-slit (Fresnel) diffraction. However, the Fraunhofer diffraction was still present, and its effects were taken into account in the background measurements.

The detector C was fixed in front of the source S2; the detectors A and B were placed on a common vertical panel. Let us highlight the role played by the three detectors. The detector C destroyed the eigenstates of the photons emitted by S2. The detector B ensured that no photon passed through the aperture F2. Finally, the detector A measured the photon signal from the source S1. In summary, the detectors B and C played a controlling role and ensured that no spurious and instrumental effects could be mistaken for the anomalous effect, which had to be revealed on the detector A. The design of the box and the measurement procedure were conceived so that the detector A was not influenced by the source S2 according to the known and commonly accepted laws of physics governing electromagnetic phenomena: classical and/or quantum electrodynamics. In other words, with regards to the detector A, all went as if the source S2 would not be there at all or would be kept turned off all the time. In essence, the experiments just consisted in the measurement of the signal of the detector A (aligned with the source S1) in two different states of source lighting. Precisely, a single measurement on the detector A consisted of two steps: **(1)** Sampling of the signal on A with source S1 switched on and source S2 off; **(2)** sampling of the signal on A with both sources S1 and S2 switched on. Analogous measurements were taken on the detectors B and C. A possible non-zero difference $\Delta A = A(S1 \text{ on } S2 \text{ off}) - A(S1 \text{ on } S2 \text{ on})$ in the signal measured by A when source S2 was off or on (and the signal in B was strictly null) has to be considered as evidence for the searched anomalous effect. Following all previous discussions, let us explicitly notice once again that the geometry of the box was strongly critical in order to reveal the anomalous photon behaviour.

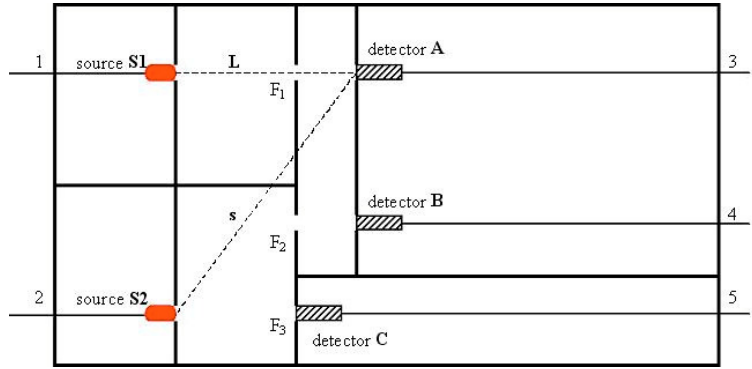

**Figure1:** Lay-out of the experimental apparatus

## 2.2. The first two experiments

The main difference between the first two experiments was in the nature of the detectors A, B, C, which were photodiodes in the former case [9] and phototransistors in the latter [10,11]. In the first experiment, the plane containing the detectors A, B was movable (the distance was varied by steps of 1.5 cm on the whole range of 10 cm). This allowed one to study how the phenomenon changes with distance from the sources. In the second experiment, the plane was fixed at a distance of 1 cm. The outcomes of the first experiment were positive. Moreover, the phenomenon obeyed the threshold behaviour predicted by the analysis of the Cologne and Florence experiments. In particular, $\Delta A$ ranged from $(2.2 \pm 0.4)$ µV to $(2.3 \pm 0.5)$ µV, values well below the threshold energy $E_{0,em} = 4,5$ µV, and the anomalous effect was observed within a distance of at most 4 cm from the sources. The results of the second experiment confirmed those of the first one. The value of the difference measured on the detector A was $(0.008 \pm 0.003)$µV, which is consistent, within the error, with the difference measured in the first experiment, provided that the unlike efficiencies of the phototransistors with respect to those of the photodiodes are taken into account [10,11]. Furthermore, not having any indication on the LLI breakdown time evolution it was necessary to adapt the sampling time procedures of the signal on the detectors to the search for the anomalous effects. In order to catch these effects we used two different time procedures to sample the signals on the detectors in the two experiments. We indeed realised that the sampling time procedure was apparently crucial in this sense and related to the type of detector. In table 1 we report the results of the second experiment which will be useful in a following discussion.

## 2.3. The third experiment

In order to test the apparent bond between detectors and sampling time procedures, the experiment was performed by means of the box with photodiodes, but using the sampling-time procedure adopted with phototransistors. The results of this third experiment were consistent with those of the two previous ones. On account of the sampling procedure adopted – it turns out that there is not such a tight bond between detector and time procedure (although, as already stressed, the latter plays a very important role in giving evidence to the effect). A first global picture of these three experiments teaches us that beyond an energy threshold and a spatial threshold, a very peculiar time procedure in sampling the signals is needed in order to track and detect LLI breakdown. The third experiment was repeated several times over a whole period of four months, in order to collect a fairly large amount of samples and hence have a significant statistical reproducibility of the results. Thanks to this large quantity of data, it was possible to study the distribution of the differences of signals on the detector A. We considered two types of differences: the differences containing the anomalous signal ΔA = A (S1 on S2 off)-A (S1 on S2 on) and the blank differences ΔA' = A (S1 on S2 on)-A (S1 on S2 on). In Fig.2 we show that the second type of differences are all compatible with zero (inside the interval [-1; 1] μV). In Fig.3 and Fig.4, conversely, we show the differences of the first type compatible with zero and those non compatible with zero respectively. These data are apparently at variance with electrodynamics which expects that all of the differences of the two types should be compatible with zero. This should be the case because, by the actual design of the experimental box, the detector A should not be affected by the state of lighting of the source S2. ΔA' is indeed compatible with the prediction of electrodynamics. The situation is quite different for the differences ΔA. Actually, only some values (73/180) are inside the interval [-1; 1] μV, of compatibility with zero, while most of them (107/180) lie outside of it. As it is clear from Fig.4, the differences A(S1 on, S2 off) - A(S1 on, S2 on) are positive and shifted upwards (needless to say, the stability of power supplies was constantly checked) which means that, despite the greater number of photons in the box when both sources are on, the detector A sees less photons than those seen when only S1 is on. Let's add that it is impossible to account for this systematic effect by a destructive interference between photons from the two sources, because the LEDs are incoherent sources of light. We performed statistical analysis [13] of the data and found out that there is no compatibility between the ΔA' set and the ΔA set non compatible with zero as shown in Fig.5. In particular the mean values of the two gaussian distributions are 3.81σ apart.

**Table 1:** Results of the second experiment

| | TOTAL MEAN $\overline{V}$ (μV) | | | | | | DIFFERENCES BETWEEN TOTAL MEANS | |
|---|---|---|---|---|---|---|---|---|
| Sources | $S_1$ | $S_2$ | $S_1$ | $S_2$ | $S_1$ | $S_2$ | | |
| State | ON | ON | ON | OFF | OFF | ON | SIGN | ABSOLUTE VALUE $\Delta\overline{V}$ (μV) |
| Detector A | 8.634 | | 8.626 | | | | (ON ON) > (ON OFF) | 0.008±0.003 |
| Detector B | 0.020 | | | | | 0.275 | (ON ON) < (OFF ON) | 0.255±0.003 |
| Detector C | 16.226 | | | | | 16.476 | (ON ON) < (OFF ON) | 0.250±0.003 |

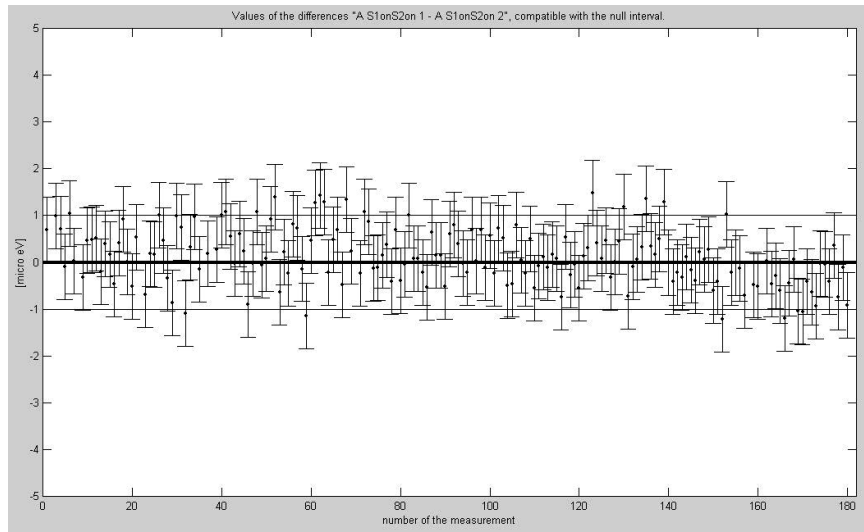

**Figure 2:** Differences ΔA' of the signal measured by the detector A with both sources on. All of them are inside the interval [-1; 1] μV.

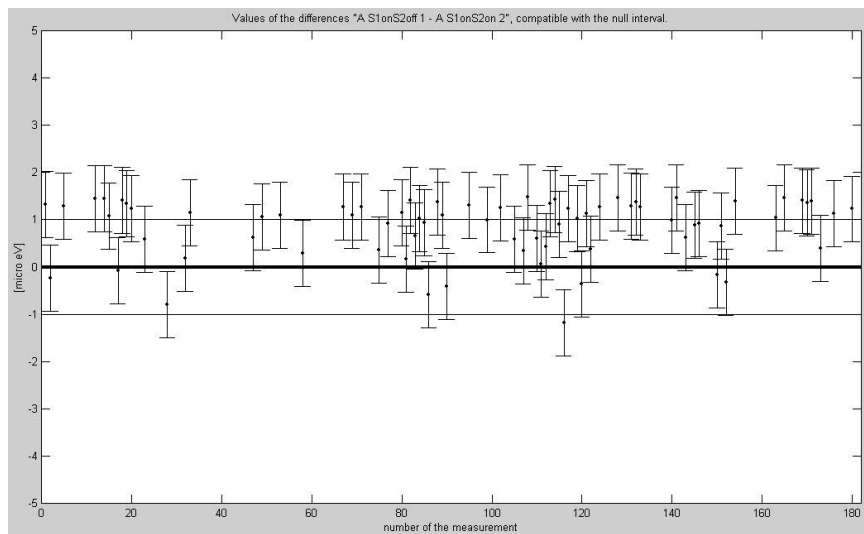

**Figure 3:** Differences ΔA, compatible with zero, of the signal measured by the detector A for the two lighting states of the sources (S1 on, S2 off) and (S1 on, S2 on). These differences lie inside the interval [-1; 1] μV.

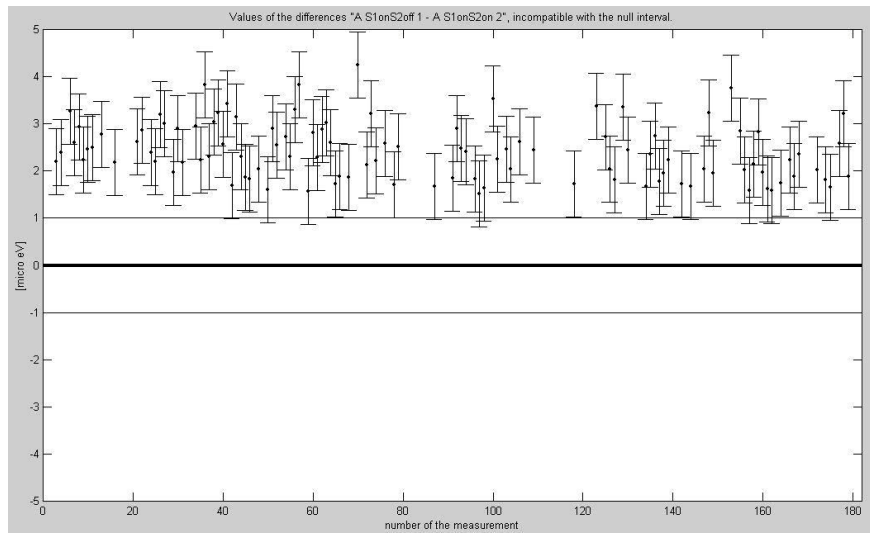

**Figure 4:** Differences ΔA, non compatible with zero, of the signal sampled on the detector A for the two lighting states of the sources (S1 on, S2 off) and (S1 on, S2 on). These differences lie outside the interval [-1; 1] µV.

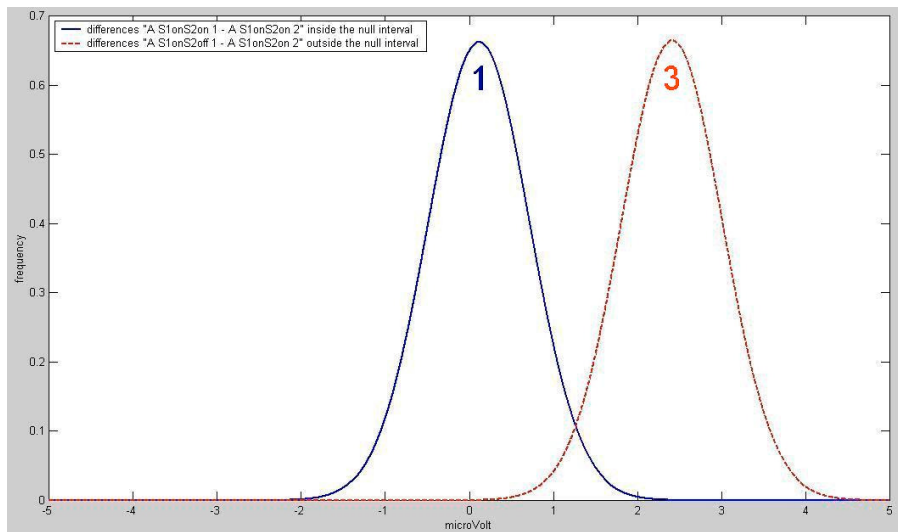

**Figure 5:** Gaussian curves for the differences ΔA' compatible with zero (blue, solid line) and ΔA non compatible with zero (red, dashed line). It is $\mu_{\Delta A'} = 0.1159$ µV, $\mu_{\Delta A} = 2.4110$ µV, $\sigma_{\Delta A'} = 0.6024$ µV, $\sigma_{\Delta A} = 0.6006$ µV, $\Delta\mu_{\Delta A', \Delta A} = 3.81\ \sigma_1$.

Before moving on to the interpretations and conclusions, we would like give some information about the temperature controls that were carried out in order to be able to exclude possible artefacts due to heat. In all of the three experiments the temperature was monitored outside the box (temperature of the laboratory) and inside the box. The former was carried on for all the sessions of measurement, while the latter was carried out only during the preliminary tests of the equipment, since we did not want to alter the geometry of the interior of the box by the presence of the thermometer. It turned out that over eight hours, the time needed to complete a session of measurements, the variation of the temperature inside the box was in the range of 0.1 – 0.3 °C. This evidence means that this variation has to be spread over an interval of time of 8 hours and in this sense, the possible variation of temperature over a period of six seconds, which is the interval of time when the LEDs are turned on, is so small that the drift with temperature both of the output offset voltage and of the amplifier gain can be neglected . In fact, they would produce a change of the measured values which is three orders and one order of magnitude lower than the values obtained in the first and third experiments (photodiode) and the second one (phototransistor) respectively. The sampling procedure was in fact designed in order to compare, by difference, values that had been sampled consecutively in order to minimise this possible source of artefact.

## 3. Interpretations and conclusions

### 3.1. Violation of Electrodynamics and LLI breakdown

It has been said that the design of the experimental set-up was carried out in order to prevent the photons of the source S2 from affecting the detector A. In other words according to the electrodynamics, one would expect the detector A not to be influenced by the lighting state of S2 as, of course, it was experimentally ascertained by keeping S1 turned off and by comparing the distributions of the dark signal on A when S2 was off with that when it was on. However, the results shown in Fig. 4 and Fig. 5 present a different situation. The turning on of S2, when S1 was already on, made the signal on A systematically change, decrease in particular, despite the higher number of photons in the box due to the lighting of two incoherent sources.
This behaviour is manifestly a violation of what the electrodynamics would predict. This is the first of the two conditions that we need to state that LLI is broken. The second condition wants the reference frame, where this experiment was performed, to be a good approximation of a local reference frame. The small dimensions of the box and the planar geometry of the experimental apparatus authorise us to state that this condition is fulfilled. As to the definition of LLI[3], these two conditions imply a local violation of the Lorentz symmetry. Since the predictions of the phenomenological theory, on which the experiment was designed, were obtained by applying a parametric deformed Minkowski metric tensor (deformed spacetime) to the outcomes of superluminal propagation experiments where LLI is broken, we interpret the effects, presented in Fig.4 and Fig.5 as brought about by a locally non Lorentz invariant deformed spacetime. In particular we can imagine that the energy of the photons emitted by S2 locally deforms spacetime and that this deformation expands through aperture F2 and reaches the photons emitted by S1 and steers (pilots) their propagation[4] before they are detected by A. In short we can say that this interpretation looks at the detected anomalous effects from the perspective of LLI breakdown and more precisely of a locally deformed spacetime. Let's now change the point of view and put forward a further possible interpretation in terms of Bohmian mechanics and pilot wave.

---
[3] The local Lorentz invariance is part of the Einstein equivalence principle and can be phrased as follows: the outcomes of any local non gravitational experiment is independent of the velocity of the free falling reference frame in which it is performed.

### 3.2. Bohmian mechanics and pilot wave

First of all let's see what Quantum Mechanics (Copenhagen interpretation) can say about the experimental set-up and the results obtained. The central part of Quantum Mechanics is the wave function, along with its probabilistic interpretation and its instantaneous collapse in one of the eigenstates as the quantum system is measured. The photons emitted by S2 are detected and hence measured by the detector C. In this sense the wave function, that describes the quantum system of these photons and that, before the measurement, represents the probability to find them in any place of the universe[5] (in the sense of Feynman path integrals), collapses in C and reduces to zero the probability to find them in any other point of the box. This quantum mechanical prediction is against the results of the experiments that indicate that the turning on of the source S2 influences both the reading of the detector C, of course, and that of the detector A although, by design, there should not be any correlation between S2 and A. The quantum mechanical descriptions is incomplete, at least in this case, since the results display before us a more complex behaviour of the photons that is unaccountable in terms of wave function and probability.

Let's see whether Bohmian mechanics can give a better account of the results in terms of the realistic interpretation of the wave function as a pilot wave. According to Bohmian mechanics, the hollow pilot wave of a quantum entity pervades all the universe in a non local sense, accompanies and steers the motion of the corpuscle associated with it and of other corpuscles of the same kind. If we refer this picture to the experiments we can say that the pilot waves of the photons emitted by S2 expand to regions of space optically forbidden to the associated photons, pass through the aperture F2 and interact with the photons emitted by S1. Consequently, the change ΔA in the A signal - in the absence of any change in the signal of the detectors B and C - finds a natural explanation, in Bohmian quantum mechanics, in terms of the interaction of the S1 photons (and their pilot waves) with the pilot waves (of the S2 photons) passed through F2. The role played by the aperture F2 is fundamental, since, although pilot waves can penetrate in optically forbidden regions, nonetheless the mass distribution and density are expected to affect their propagation. Hence, they can only pass through space regions with a lower mass density. If this interpretation is correct, the double-slit box experiments do provide for the first time, among the others, direct evidence for the Einstein, de Broglie, Bohm waves.

---

[4] Just like the curved spacetime around the Sun curve the trajectory of the photons from a distant star.
[5] Or, at least, of the entire box

### 3.3. Failure of Bohr's Priciple of Complementarity

A careful analysis seemingly shows that the outcomes of the box experiments, in particular those of the second one, invalidate the Bohr principle of complementarity (BCP) too. Let us explain why. In essence, BCP when applied to light, states that it is impossible to put in evidence both the wave (interference fringes) and the particle behaviour (which way information) of a photon in the same experiment. We will show that this is not the case in the box experiments. As it was extensively described in [11] the box was designed according to the laws of geometrical optics in order to exploit the corpuscular nature of the photons (refer to [11] for details). However if we refer to table 1, where the results of the second experiment are reported, it is possible to envisage a further interpretation in terms of waves and interference fringes. Let's refer to Table 1 and compare for each detector how the signal changes between the two lighting conditions. The signal on C remained constant (within the device working fluctuations). Things are different for the detectors A and B. Let's refer, first, to the detector B. As the lighting condition changes from one source turned on (S1off_S2on) to two sources on (S1on_S2on), the value of the signal on B decreases of one order of magnitude. Referring to Fig.6, one can say that when (S1off_S2on) the detector B sees a certain density of photons and when both sources are on (S1on_S2on) this density decreases. In other words, we could say that, everything goes as if, when both sources are on, it formed a dark fringe where the detector B is. The same kind of considerations can be done for the detector A. When the lighting condition changes from (S1on_S2off) to (S1on_S2on) the signal on A increases a little bit, although this cannot be imputed to the passage of photons from F2 since the detector B is always underneath the maximum dark voltage established experimentally. Referring again to Fig.6, one can say that when (S1on_S2off) the detector A sees a certain density of photons and when both sources are on (S1on_S2on) this density increases. In other words, we could say that, everything goes as if, when both sources are on, it formed a bright fringe where the detector A is.
This anomalous behaviour of the detectors A and B can still be explained on the basis of Bohmian quantum mechanics, by interpreting it as manifestations of an anomalous interference phenomenon involving pilot waves. In this view, the anomalous signals at the detectors A and B can be regarded as part of an interference pattern. Consequently, the results obtained in these experiments seemingly invalidate Bohr's principle of complementarity. Let us mention that compatible results were obtained by Afshar [14] in a double-slit experiment with photons, where both the interference pattern and the which-way information are known.

### 3.4. Pilot wave and spacetime deformation

In the previous sections we discussed the results of the experiments and showed that two main conclusions can be drawn: the breakdown of LLI interpreted in terms of local spacetime deformation that steers the propagation of photons and the presence of the pilot wave along with its piloting effect of photons. The similarity of the descriptions of the effects either in terms of deformed spacetime or pilot waves led us to put forward the intriguing hypothesis of a possible connection between these two pictures, namely between the quantum wave (in the interpretation by de Broglie and Bohm, as a pilot wave) and the breakdown of local Lorentz invariance, described by the formalism of Deformed Special Relativity [5,6] (i.e., in terms of a modified Minkowski metric). In particular we state that what we call pilot wave is nothing but a deformation of the local space-time geometry, intimately bound to the quantum entity considered (photon) which, in this sense, becomes a more complex object than that described by the quantum electrodynamics and even by Bohmian mechanics. In particular, with regards to the photon we can say that most of its energy is concentrated in a tiny extent (complying with electrodynamics, relativity and Minkowski space-time) and the rest of the energy is used to deform space-time surrounding it (violating electrodynamics, not complying with relativity and hence possessing real non-local and superluminal features). This second part of the energy is stored in the local deformation of space-time just as the Riemann curvature of space-time in General Relativity possesses its own energy momentum pseudo-tensor. Therefore, in this view, the difference of signal measured by the detector A in all of the three experiments can be interpreted as the energy absorbed by the space-time deformation itself, which cannot be detected by the central detector B. In other words, the experimental apparatus and procedure, used in these experiments, "weighed" the energy corresponding to the space-time deformation by the measured difference on the first detector. However we have to say that, after all, it is known that LLI breakdown and quantum processes are not too far from eachother especially if the latter are considered from the point of view of Bohmian mechanics [19]. We said that LLI breakdown implies superluminality which, in turn, is strongly interrelated to non-locality and, as is well known, quantum theory[6] is non-local as proved by the famous Einstein-Podolski-Rosen (EPR) effect (experimentally checked by Aspect [16] and then by Alley [17]). However, we reckon that the results of our experiments and the interpretations that we put forward, add three further important elements to the discussion on the foundations of quantum mechanics that, to the best of our knowledge, have not been considered yet. All of them are based on the very general perspective on LLI breakdown in term of local space-time deformation. The first element is the hypothesis that the pilot wave (the real counterpart of the probabilistic wave function) is only a synoptic concept that hides a more complex reality for the quantum object photon and its possible manifestations, corpuscle or wave. The results of the experiments show that photons possess the well known Lorentz invariant part and a further non Lorentz invariant one which brings about the effects usually described by pilot waves such as interference fringes and non local effects. However, while pilot waves have the problem to be hollow waves which has always been a critical point in having Bohmian mechanics accepted, the deformed space-time stores in its deformation the energy which affects the propagation of photons and which is contained in the difference of signals detected by A. In this sense according to our interpretation, the box (double-slit) experiments not only provide direct evidence for the Einstein-de Broglie-Bohm waves by their effect on photon propagation, but also yield a measurement of the energy associated to them and indicate the space-time deformation as the physical entity hidden behind their synopsis. If this hypothesis is correct, pilot waves are no more hollow waves and there is no more "spooky" action at a distance and even the concept of action at a distance, which implies infinite causal speed, will have to be carefully re-examined. The second two elements that our experiments and interpretations add to the foundations of quantum mechanics have an experimental character and, hence, contain precious indications to guide in performing further experiments and to interpret the existing results (maybe negative) of past experiments. It has been said that Bohmian mechanics implies non-locality and, hence, the breakdown of LLI. The second element has to do with the requirements to fulfil in order to detect the effects due to LLI breakdown: the existence of an energy threshold and a space threshold for the locally non Lorentz invariant effects to show up. The third element is more subtle and has to do with time and in particular, with the time evolution of LLI breakdown and hence of the effects brought about by the space-time deformation or pilot wave. Since space-time is deformed the geometry of space is no more homogeneous and isotropic and the geometry of time is no more homogeneous. As to the time evolution of the effects due to LLI breakdown, this means that the experimenter will have to adapt the time evolution of his measuring equipment, which is a LLI compliant time, and hence homogeneous, to a completely a priori unknown evolution which will have to be studied and followed experimentally.

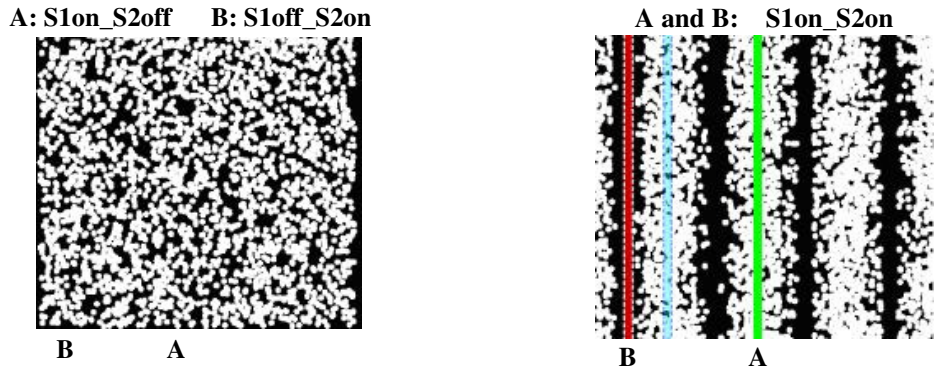

Fig. 6 Sketch of the wave behaviour of photons in the second experiment. When one source is on (left) the detectors A and B detect a certain density of photons corresponding to the signal from S1 and the dark respectively. When both sources are on (right), A sees a higher density of photons (bright fringe) while B sees a lower density (dark fringe).

---

[6] The Copenhagen interpretation of the wave function hides the non-locality in the positivistic point of view that something is real only when it is measured. Conversely, Bohmian mechanics makes non-

**3.5.    Perspectives**

---

locality explicit since it looks at the wave function as a real physical entity, although it cannot explain its apparent lack of any energy transport.

The change of the number of photons detected by A can be read as well in terms of the modification of the photon-photon cross section due to the deformed space-time associated to each photon. With this last picture of modified cross section, we report that compatible results with ours were obtained in crossed photon-beam experiments both in the microwave range [20,21] and with a $CO_2$ laser [9,13]. Crossed laser beams are certainly a much simpler experimental set-up, however, it goes without saying that, despite the apparent simple and very common appearance of the equipment, the features of LLI breakdown are so peculiar, as we have extensively pointed out, that require subtle care in tuning all the experimental features in order to make out the anomalous effect due to deformed space-time. However the intersecting of laser beam is a very suitable set-up to deepen the study of different features: spatial extension by varying horizontally and vertically the crossing region of the beams; timing of the deformation by applying different chopper frequencies and sampling time procedures; investigating the effects of deformed space-time on the frequency of photons; attempts to work as close as possible to the single photon; drawing interesting similarities between the non-linear and non-local effects studied by non linear optics in nonlinear media (liquid crystals) and non linear non local effects of space-time; visualize by CCD the deformation of the laser beam spot which is the unmistakable evidence of deformed spacetime through which photons propagate as already tried and reported in [9].

## References

[1] L. de Broglie, La réinterpretation de la Mécanique Ondulatoire, Gauthier–Villars, Paris, 1971; D. Bohm, Causality and Chance in Modern Physics, with a foreword by L. de Broglie, London, 1957; J.S. Bell, Speakable and Unspeakable in Quantum Mechanics, Cambridge Univ. Press, Cambridge, 1987, and references therein.

[2] L. de Broglie, J. Andrade y Silva, Phys. Rev. 172 (1968) 1284.

[3] C.M. Will, Theory and Experiment in Gravitational Physics, Cambridge Univ. Press, 1993, and references therein.

[4] A. Kostelecky (Ed.), CPT and Lorentz Symmetry, vols. I and II, World Scientific, Singapore, 1999 and 2002, and references therein.

[5] F. Cardone, R. Mignani: *Energy and Geometry – An Introduction to Deformed Special Relativity* (World Scientific, Singapore, 2004).

[6] F. Cardone, R. Mignani: *Deformed Spacetime – Generalizing Interactions in Four and Five Dimensions* (Springer, Dordrecht, 2007).

[7] A. Enders, G. Nimtz, Phys. Rev. E 48 (1993) 632.

[8] A. Ranfagni, P. Fabeni, G.P. Pazzi, D.Mugnai, Phys. Rev. E 48 (1993) 1453.

[9] F. Cardone, R. Mignani, W. Perconti and R. Scrimaglio, Phys. Lett. A 326, 1 (2004).

[10] F. Cardone, R. Mignani, W. Perconti, A. Petrucci and R. Scrimaglio, Int. J. Mod. Phys. B 20, 85 (2006).

[11] F. Cardone, R. Mignani, W. Perconti, A. Petrucci and R. Scrimaglio, Int. J. Mod. Phys. B 20, 1107 (2006).

[12] F. Cardone, R. Mignani, W. Perconti, A. Petrucci and R. Scrimaglio, Annales de la Fondation Louis de Broglie, Vol.33, no.3, (2008)
[13] F. Cardone, R. Mignani, Int. J. Mod. Phys. B Vol.21, no.26, 4437-4471 (2007)
[14] S. S. Afshar, Proc. SPIE 5866, 229 (2005).
[15] L. Kostro, Einstein and the Aether (Apeiron, Montreal, 2000).
[16] A. Aspect, P. Granger and G. Roger, Phys. Rev. Lett. 49, 91 (1982).
[17] C. O. Alley and Y.-H. Shih, Phys. Rev. Lett. 61, 2921 (1988).
[18] L. de Broglie, La Reinterpretation de la Mecanique Ondulatoire (Gauthier-Villars, Paris, 1971).
[19] Goldstein, Bohmian Mechanics, The Stanford Encyclopedia of Philosophy (Summer 2006 Edition), ed. E. N. Zalta, http://plato.stanford.edu/archives/sum2006/ entries/qm-bohm.
[20] A. Ranfagni, D. Mugnai and R. Ruggeri, Phys. Rev. E 69, 027601 (2004).
[21] A. Ranfagni and D. Mugnai, Phys. Lett. A 322, 146 (2004).